\newif\ifpictures
\picturestrue
\newif\iflong
\longfalse

\documentclass[12pt]{amsart}
\usepackage{amssymb}
\usepackage[dvips]{graphicx}

\ifpictures
\usepackage{psfrag}
\fi

\numberwithin{equation}{section}

\newtheorem{thm}{Theorem}
\newtheorem{prop}[thm]{Proposition}
\newtheorem{lemma}[thm]{Lemma}
\newtheorem{cor}[thm]{Corollary}

\theoremstyle{definition}

\newtheorem{example}[thm]{Example}
\newtheorem{remark1}[thm]{Remark}
\newtheorem{openproblem1}[thm]{Open problem}
\newtheorem{definition}[thm]{Definition}

\newenvironment{rem}{\begin{remark1}\rm}{\end{remark1}}

\newenvironment{openproblem}{\begin{openproblem1}\rm}{\end{openproblem1}}

\numberwithin{thm}{section}

\newcounter{FNC}[page]
\def\newfootnote#1{{\addtocounter{FNC}{2}$^\fnsymbol{FNC}$%
     \let\thefootnote\relax\footnotetext{$^\fnsymbol{FNC}$#1}}}

\newcommand{\N}{\mathbb{N}}

\newcommand{\R}{\mathbb{R}}

\newcommand{\Z}{\mathbb{Z}}

\title[Games of fixed rank]{
  Games of fixed rank: \\ A hierarchy of bimatrix games}

\author{Ravi Kannan}

\address{R.~Kannan: Dept.\ of Computer Science,
  Yale University, P.O.\ Box 208285, New Haven, CT 06520--8285, USA}

\email{kannan@cs.yale.edu}

\author{Thorsten Theobald}

\address{T.~Theobald: Institut f\"ur Mathematik, MA 6--2,
  Technische Universit\"at Berlin, 
  Stra{\ss}e des 17.~Juni 136, D-10623 Berlin, Germany}

\email{theobald@math.tu-berlin.de}

\thanks{Part of this work was done while the second author was
a Feodor Lynen fellow of the Alexander von Humboldt Foundation
at Yale University.}

\begin{document}

\begin{abstract} 
We propose a new hierarchical approach to understand the 
complexity of the open problem of computing a Nash equilibrium in
a bimatrix game. Specifically,
we investigate a hierarchy of bimatrix games $(A,B)$ which
results from restricting the rank of the matrix $A+B$
to be of fixed rank at most $k$. For every fixed $k$, this class
strictly generalizes the class of zero-sum games, but is a very special
case of general bimatrix games.
We show that even for $k=1$ the set of Nash equilibria of these
games can consist of an arbitrarily large number of connected components.
While the question of exact polynomial time algorithms to find
a Nash equilibrium remains open for games of fixed rank,
we can provide polynomial time algorithms for finding an 
$\varepsilon$-approximation.
\end{abstract}

\maketitle




\section{Introduction}

Models of non-cooperative game theory serve to analyze situations
of strategic interactions. Driven by current developments in
auction theory as well as in equilibria models for the internet,
the basic model of a \emph{Nash equilibrium} has recently attracted much
attention (see for example 
the survey by Papadimitriou \cite{papadimitriou-2001} or the recent papers
\cite{bvv-2005,dgp-2005,papadimitriou-roughgarden-2005,
savani-stengel-2004}).

In \cite{von-neumann-morgenstern-44}, von Neumann and Morgenstern introduced
the model 
of \emph{zero-sum games}, 
which are described by a single $m \times n$-matrix $A$.
These games always possess an equilibrium, and the set of all equilibria
(which is a polyhedral set and thus in particular connected) 
can be computed efficiently using linear
programming (see, e.g., \cite{dantzig-b51}).

Nash investigated the model of \emph{bimatrix games} $(A,B)$ 
(and more generally $N$-player games) 
\cite{nash-pams-50,nash-annals-51}, in which the gain 
of one player does not necessarily agree with the loss of the other player,
thus adding much expressive power to the model of zero-sum games. 
By Nash's results any bimatrix game has at least one equilibrium.
However, it is still not known whether an equilibrium can
be computed in polynomial time, 
and that question has been named by Papadimitriou
to be the most concrete open question on the boundary of {\bf P}
\cite{papadimitriou-2001}.
Even approximability in polynomial time is not known; for 
quasi-polynomial time approximation algorithms 
see Lipton, Markakis, and
Mehtat \cite{lmm-03}.

Thus, it will be of interest to impose restrictions on bimatrix games
which while preserving expressive power of the games may admit simple
polynomial time algorithms.
Recently, Lipton et al. \cite{lmm-03} investigated games where both
payoff matrices $A,B$ are of fixed rank $k$. They showed that in
this restricted model a Nash equilibrium can be found in polynomial time.
However, for
a fixed rank $k$, the expressive power of that model is limited; in
particular, most zero-sum games do not belong to that class.

In this paper, we propose and investigate a related model
based on low-rank restrictions, but which is a strict superset of
the model of zero-sum games.
The viewpoint we start with is that in a zero-sum game, the sum
of the payoff matrices $C:=A+B \in \R^{m \times n}$ 
is the zero matrix, which for our purposes
we consider as a matrix of rank~0. In a general bimatrix game the
rank of $C$ can take any value up to $\min \{m,n\}$. Here, we consider
the hierarchy given by the class of games in which we restrict 
$C$ to be of rank at most $k$ for some given $k$. We call these
games \emph{rank $k$-games}.

\medskip

\noindent
{\bf Our contributions.} We show that the
expressive power of fixed rank-games is significantly larger than that of
zero-sum games.
In order to provide this separation, we exhibit a sequence of 
$d \times d$-games of rank 1 whose number of connected components
of equilibria exceeds any given constant. Our lower bound for the 
maximal number
of Nash equilibria of a $d \times d$-game is linear in $d$.
This bound is not tight.

Although the problem of finding a Nash equilibrium in a game of
fixed rank is a very special case of the problem of finding 
a Nash equilibrium in an arbitrary bimatrix game, we do not know
if there exists an exact polynomial time algorithm for this problem.
Note that the problem strictly generalizes linear programming
(see, e.g., \cite[Ch.~13.2]{dantzig-b51} for the equivalence of linear
programming and zero-sum games).

However, we provide approximation results for two approximation
models. Firstly, we propose a model of $\varepsilon$-approximation
for rank $k$-games.
Using existing results from quadratic optimization, we show that 
we can approximate Nash equilibria of constant rank-games in polynomial
time, with an error relative to a natural upper bound on the
``maximum loss'' of the game (as defined in Section~\ref{se:approximation}).

Secondly, we present a polynomial time approximation algorithm for 
\emph{relative} approximation
(with respect to the payoffs in a Nash equilibrium)
provided that the matrix $C$ has a nonnegative decomposition.

\section{Preliminaries}

We consider an $m \times n$-bimatrix game with payoff matrices 
$A,B \in \Z^{m \times n}$.
Let 
\[
  \mathcal{S}_1 = \big\{ x \in \R^m \, : \, \sum_{i=1}^m x_i = 1 \, , \: 
  x \ge 0 \big\} \;  \text{ ~~and~~ } \; 
  \mathcal{S}_2 = \big\{ y \in \R^n \, : \, \sum_{j=1}^n y_j = 1 \, , \:
  y \ge 0 \big\}
\]
be the sets of mixed strategies of the two players, and let
$\overline{\mathcal{S}}_1 = \{x \in \R^m \, : \, \sum_{i=1}^m x_i = 1 \}$
and 
$\overline{\mathcal{S}}_2 = \{y \in \R^n \, : \, \sum_{j=1}^n y_j = 1 \}$
denote the underlying linear subspaces. The first player (the row player)
plays $x \in  \mathcal{S}_1$ and the second player (the column player)
plays $y \in \mathcal{S}_2$. The payoffs for player~1 and player~2 are
$x^T A y$ and $x^T B y$, respectively.

Let $C^{(i)}$ denote the $i$-th row of a matrix $C$ (as a row vector), 
and let $C_{(j)}$ denote the $j$-th column of $C$ (as a column vector).
A pair of mixed strategies $(\overline{x}, \overline{y})$
is a \emph{Nash equilibrium} if
\begin{equation}
\label{eq:defnash}
  \overline{x}^T A \overline{y} \ \ge \ x^T A \overline{y}
  \quad \text{ and } \quad \overline{x}^T B \overline{y} \ \ge \ \overline{x}^T B y
\end{equation}
for all mixed strategies $x$, $y$.
Equivalently, $(\overline{x},\overline{y})$ is a Nash equilibrium 
if and only if
\begin{equation}
\label{eq:bestresponse1}
  \overline{x}^T A \overline{y} \ = \ 
  \max_{1 \le i \le m} A^{(i)} \overline{y}
  \quad \text{ and } \quad
  \overline{x}^T B \overline{y} \ = \ 
  \max_{1 \le j \le n} \overline{x}^T B_{(j)} \, .
\end{equation}

\subsection{Economic interpretation of low-rank games.}

If $A + B = 0$ then the game is called a zero-sum game. The economic
interpretation of a zero-sum game is ``What is good for player~1 is
bad for player~2''. In order to describe game-theoretic
situations which are close to that behavior, we consider a 
model where $a_{ij} + b_{ij}$ is a function which depends only on
$i$ and $j$
\[
  a_{ij} \ + \ b_{ij} \ = \ f(i,j)
\]
and where $f$ is a simple function. If 
$f: \{1,\ldots, m\} \times \{1, \ldots, n\} \to \Z$ 
is an additive function, $f(i,j) = u_i + v_j$ with constants
$u_1, \ldots, u_m, v_1, \ldots, v_n$, then there
is an \emph{equivalent} zero-sum game, i.e., a game having the
same set of Nash equilibria. Namely, define the payoff matrices
$A'$ and $B'$ by
\[
  a'_{ij} \ = \ a_{ij} - v_j \, , \quad
  b'_{ij} \ = \ b_{ij} - u_i \, .
\]
That is, $A'$ results from $A$ by adding the column vector 
$(v_j, \ldots, v_j)^T$ to the $j$-th
column $(1 \le j \le n)$ and 
$B'$ results from $B$ by adding the row vector
$(u_i, \ldots, u_i)$ to the $i$-th row $(1 \le i \le m)$.
Now
\[
  \overline{x}^T A' \overline{y} - x^T A' \overline{y} \ = \
  \overline{x}^T A \overline{y} - \sum_{j=1}^n v_j \overline{y}_j
  - x^T A \overline{y} + \sum_{j=1}^n v_j \overline{y}_j
  \ = \ \overline{x}^T A \overline{y} - x^T A \overline{y}
\]
and a similar relation w.r.t.\ $B$ holds. So
the zero-sum game $(A',B')$ has the same Nash equilibria as $(A,B)$.
We remark that the case $v_j = 0$ yields the row-constant games
introduced in \cite{isaacson-millham-80}.

If $f$ is a multiplication function, $f(i,j) = u_i v_j$ 
with constants
$u_1, \ldots, u_m, v_1, \ldots, v_n$, this is a rank 1-game.
If $f$ is a sum of $k$ multiplication functions, this
is a game of rank at most $k$.

Rank-1 games also occur under the term ``multiplication games'' in
the paper \cite{bulow-levin-2003} by Bulow and Levin.

\subsection{Approximate Nash equilibria}

We also consider approximate equilibria. To define them,
suppose $x$ is not necessarily an optimal strategy for player~1
given that player~2 has played $y$. Then the ``loss'' for player~1
(from optimum) is $\max_i A^{(i)} y - x^T A y$. Similarly, if $y$ is
not optimal for player~2 given that the first player has played $x$,
the loss for player~2 would be $\max_j x^T B_{(j)} - x^T B y$. We will use the
total of these two losses -- i.e.,
\[
\ell(x,y) \ = \ \max_i A^{(i)} y + \max_j x^T B_{(j)} - x^T (A+B) y
\]
as a measure of how much $(x,y)$ is off from equilibrium.
For a matrix $X \in \R^{m \times n}$ let
$|X| = \max_{1 \le i \le m, 1 \le j \le n} |x_{ij}|$.

\begin{definition} \label{de:approxnasheq}
For $\varepsilon \ge 0$, a pair $(x,y)$ of mixed strategies 
is an \emph{$\varepsilon$-approximate equilibrium} if 
\begin{equation}
  \label{eq:approx1}
  \ell(x,y) \ \le \ \varepsilon |A+B| \, . 
\end{equation}
\end{definition}

Note that the term $|A+B|$ on the right hand side
provides a stronger approximation model 
compared to the term $|A|+|B|$. Also
observe that $|A+B|$ is an upper bound for the term $x^T (A+B) y$.
For a game with $A - B \neq 0$,
a pair of strategies is an exact equilibrium if and only if it
is a 0-approximate equilibrium. Besides the notion of 
``absolute'' approximation in Definition~\ref{de:approxnasheq}, 
in Section~\ref{se:relativeapprox}
we will also consider a notion of ``relative'' approximation.

\begin{lemma}
Suppose $(\overline{x},\overline{y})$ is 
an $\varepsilon$-approximate equilibrium. Then
\begin{equation}
  \label{le:epsilonapprox1}
  x^T A \overline{y} + \overline{x}^T B y - \overline{x}^T (A+B) \overline{y}
  \ \le \ \varepsilon |A+B| \text{ for any other 
  mixed strategies $x, y\,.$} 
\end{equation}
Also, conversely, if a pair of mixed strategies $(\overline{x},\overline{y})$ 
satisfies~\eqref{le:epsilonapprox1} then it is an 
$\varepsilon$-approximate equilibrium.
\end{lemma}

\begin{proof} The proof follows from the equivalence of the 
statements~\eqref{eq:defnash} and~\eqref{eq:bestresponse1}.
\end{proof}

\subsection{Approximation of games by low rank games} 
If the matrix $C = A+B$ of a bimatrix game is ``close'' to a 
game with rank $k$, then the game can be approximated by a 
rank $k$-game $(A',B')$ in such a way that the Nash equilibria of the 
original game $(A,B)$ remain approximate Nash equilibria in the game $(A',B')$.

\begin{definition}
Let $(A,B)$ be an $(m \times n)$-game and $C=A+B$. 
If a matrix $C' \in \R^{m \times n}$ satisfies
$|C - C'| < \varepsilon (|A+B|)$ then the game $(A',B')$
with $A' = A + \frac{1}{2}(C' - C)$, 
$B' = B + \frac{1}{2}(C' - C)$ \emph{$\varepsilon$-approximates}
$(A,B)$.
\end{definition}

Note that $A' + B' = C'$.

Under the perturbation of the game, Nash equilibria of the original
game are approximate equilibria of the perturbed game:

\begin{thm}
\label{th:approxgame}
Let $(A',B')$ be an $\varepsilon$-approximation of the game $(A,B)$
and $\varepsilon < 1$.
If $(\overline{x},\overline{y})$ is a Nash equilibrium of the
game $(A,B)$, then $(\overline{x},\overline{y})$ is a
$3 \varepsilon$-approximate Nash equilibrium for the game $(A',B')$.
\end{thm}

\begin{proof}
The loss $\ell'(\overline{x},\overline{y})$ for 
$(\overline{x},\overline{y})$ with respect to the perturbed
game $(A',B')$ satisfies
\begin{eqnarray*}
  \ell'(\overline{x},\overline{y}) & \le & \varepsilon + \max_i (A' - A)^{(i)} \overline{y} + \max_j \overline{x}^T
    (B' - B)_{(j)} - \overline{x}^T (C' - C) \overline{y} \\
   & \le & \varepsilon + \frac{\varepsilon}{2} + \frac{\varepsilon}{2} + 
   \varepsilon \ = \ 3 \varepsilon
\end{eqnarray*}
\end{proof}

We can apply the Singular Value Decomposition (SVD) to approximate the 
matrix~$C$ by a matrix of some given rank $k$. The approximation factor
in Theorem~\ref{th:approxgame} is then a function of the 
singular values of $C$.

\section{The expressive power of low rank games}

\subsection{The geometry of Nash equilibria}

One measure for the expressive power of a game-theoretic model is
the number of Nash equilibria it can have (depending on the number
of strategies $m,n$). For simplicity, we will concentrate on the case
$d := m = n$.
If the Nash equilibria are not isolated, then
me might count the number of connected components, but we will
mainly concentrate on non-degenerate games in which there exist only
a finite number of Nash equilibria.

Note that the usual definition of a non-degenerate game
is slightly stronger than just requiring isolated Nash equilibria
(see the discussion in \cite{stengel-handbook-2002}).

\begin{definition} \label{de:nondegeneracy}
A bimatrix game is called \emph{non-degenerate}
if the number of the pure best responses of player~1 to a mixed
strategy $y$ of player~2 never exceeds the cardinality of the
support $\mathrm{supp~}y := \{j \, : \, y_j \neq 0\}$
and if the same holds true for the best pure responses of player~2.
\end{definition}

If $d \le 4$, then a non-degenerate $d \times d$-game can have at
most $2^d-1$ Nash equilibria, and this bound is tight 
(see \cite{keiding-97,mclennan-park-97}).
For $d \ge 5$, determining the maximal number of a non-degenerate
$d \times d$-game is an open problem (see \cite{stengel-99}).
Based on McMullen's Upper Bound Theorem for polytopes, 
Keiding \cite{keiding-97} gave an upper bound of $\Phi_{d,2d} - 1$,
where 
\[
  \Phi_{d,k} \ := \ \begin{cases}
    \frac{k}{k-\frac{d}{2}} \binom{k-\frac{d}{2}}{k-d} & \text{if $d$ even} \, , \vspace*{1ex} \\
    2 \, \binom{k-\frac{d+1}{2}}{k-d} & \text{if $d$ odd} \, .
  \end{cases}
\]
A simple class of configurations which yield an
exponential lower bound of $2^d-1$ is the game where the payoff matrices
of both players are the identity matrix $I_d$ 
(see~\cite{quint-shubik-2002}).

The best known lower bound was given by von Stengel \cite{stengel-99}, 
who showed that
for even $d$ there exists a non-degenerate $d \times d$-game having
\begin{equation} \label{eq:tau}
  \tau(d) := f(d/2) + f(d/2-1)-1
\end{equation}
  Nash equilibria, where
$
  f(n) \ := \ \sum_{k=0}^n \binom{n+k}{k} \binom{n}{k} \, .
$
Asymptotically, $\tau$ grows as
$\tau(d) \sim 0.949 \frac{(1+\sqrt{2})^d}{\sqrt{d}}$.

If the ranks of $A$ and $B$ are bounded by a fixed constant,
then the number of Nash equilibria is bounded polynomially in $d$:

\begin{thm} \label{th:abconstant}
For any $d \times d$-bimatrix game $(A,B)$ in which the ranks of both $A$ 
and $B$ are bounded by a fixed constant $k$,
the number of connected components of the
Nash equilibria is bounded by $\binom{d}{k+1}^2$.

In particular, for a non-degenerate game the number
of Nash equilibria is at most $\binom{d}{k+1}^2$, i.e., that number is
bounded polynomially in $d$.
\end{thm}

\begin{proof} Let $A$ and $B$ be of rank at most $k$.
The column space of $A y$ has dimension at most $k$.
By applying Caratheodory's Theorem on the columns of $Ay$,
it was shown in \cite[Theorem 4]{lmm-03} that for every
Nash equilibrium $(\overline{x},\overline{y})$ there exists
a Nash equilibrium $(\overline{x},y')$ in
which the second player plays at most $k+1$ pure strategies
with positive probability.
The same argument can be used to bound the number of pure
strategies which are used by player~1.
It follows from that argument that there exists a
continuous path from the original Nash equilibrium to the
Nash equilibrium with small support.

Since for a given support of the Nash equilibria, the set
of Nash equilibria with that support
is a polyhedral set, the number of connected
components of the Nash equilibria of game $(A,B)$ is at most
$\binom{d}{k+1}^2$.
\end{proof}

Now we show that the expressive power of fixed rank-games is significantly
higher than the expressive power of zero-sum games. In order to show this,
we prove that the number of Nash equilibria 
of a rank~1-game can exceed any given constant and give a linear lower bound.

\begin{thm} \label{th:linearbound}
For any $d \in \N$ there exists a non-degenerate $d \times d$-game of
rank~1 with at least $2d-1$ many Nash equilibria.
\end{thm}

The following questions remain unsolved.

\begin{openproblem}
Is the maximal number of Nash equilibria for non-degenerate
$d \times d$-games of rank $k$ smaller than the maximal number
of Nash equilibria of non-degenerate $d \times d$-games of arbitrary rank?
Is the maximal number of Nash equilibria for non-degenerate
$d \times d$-games of rank $k$ polynomially bounded in $d$?
\end{openproblem}

In order to prove Theorem~\ref{th:linearbound}, we use
the following representation of Nash equilibria introduced
by Mangasarian \cite{mangasarian-64}.

\begin{definition} \label{de:mangasarian}
For an $m \times n$-bimatrix game $(A,B)$, the 
polyhedra $\overline{P}$ and $\overline{Q}$ are defined by
\begin{eqnarray}
  \: \quad \overline{P} & = & \{(\overline{x}, v) \in \R^m \times \R \, \, : \,
\underbrace{\overline{x} \ge 0}_{\text{inequalities } 1, \ldots, m}, \;
  \underbrace{\overline{x}^T B \le {\bf 1}^T v}_{\text{inequalities } m+1, \ldots, m+n}, \; {\bf 1}^T \overline{x} = 1 \} \, \label{eq:p} \, , \\
  \: \quad \overline{Q} & = & \{(\overline{y}, u) \in \R^n \times \R \, \, : \,
  \underbrace{A \overline{y} \le {\bf 1} u}_{\text{inequalities } 1, \ldots, m}
 , \; \underbrace{\overline{y} \ge 0}_{\text{inequalities } m+1, \ldots, m+n} ,
 \; {\bf 1}^T \overline{y} = 1 \} \label{eq:q} \, .
\end{eqnarray}
\end{definition}

A pair of mixed strategies 
$(\overline{x},\overline{y}) \in \mathcal{S}_1 \times \mathcal{S}_2$ 
is a Nash equilibrium if and only if there exist
$u, v \in \R$ such that
$(\overline{x},v) \in \overline{P}$,
$(\overline{y},u) \in \overline{Q}$ and for all $i \in \{1, \ldots, m+n\}$,
the $i$-th inequality of $\overline{P}$ or $\overline{Q}$ is binding.
Here, $u$ and $v$ represent the payoffs of player~1 and player~2, respectively.
For $i \in \{1, \ldots, m\}$ we call the inequality $x_i \ge 0$ the
\emph{$i$-th nonnegativity inequality} of $P$, and for $j \in \{1, \ldots, n\}$
we call the inequality $\overline{x}^T B_{(j)} \le v$ the 
\emph{$j$-th best response inequality} of $P$. And analogously for $Q$.

\subsection{A class of low rank games with arbitrarily many Nash equilibria}

We construct a sequence $(A_d, B_d)$ 
of $d \times d$-games of rank~1 in 
which all pairs $(i,i)$ of pure strategies $(1 \le i \le d)$ are Nash
equilibria. For convenience of notation, we will omit the index $d$ 
in the notation of the game.
In order to achieve the desired properties, 
we enforce that for every
$i \in \{1, \ldots, d\}$ the element $a_{ii}$ is the maximal element
in the $i$-th column of $A$ and the element $b_{ii}$ is the maximal
element in the $i$-th row of $B$. 

Let us begin with an auxiliary sequence of games $(\overline{A}, 
\overline{B})$. 
Let $\overline{A}, \overline{B} \in \R^{d \times d}$ be defined by
\begin{equation}
  \label{eq:abarbbar}
  \overline{a}_{ij} \ = \ \overline{b}_{ij} \ = \ \ - (i-j)^2 \, .
\end{equation}
Then for every $i \in \{1, \ldots, d\}$ the element $\overline{a}_{ii}$ is the 
largest element in the $i$-th column of $\overline{A}$, and the
element $\overline{b}_{ii}$ is the largest element in the $i$-th row of $B$.
Expanding~\eqref{eq:abarbbar} shows that both $\overline{A}$ and
$\overline{B}$ can be written as the sum of three rank 1-matrices;
since $\overline{A} = \overline{B}$, it follows that the game 
$(\overline{A},\overline{B})$ is a rank $3$-game.

In order to transform $(\overline{A}, \overline{B})$ into a rank 1-game,
we observe that adding a constant column vector 
to a column of $A$ or adding a
constant row vector to a row of $B$ does not change the set of Nash equilibria.
For $j \in \{1, \ldots, d\}$, we add the 
constant vector $(2j^2, \ldots, 2j^2)^T$
to the $j$-column of $\overline{A}$, and for $i \in \{1, \ldots, d\}$ we add the
constant vector $(2i^2, \ldots, 2 i^2$) to the $i$-th row of $\overline{B}$.
Let $A, B \in \R^{n \times d}$ be the resulting matrices, i.e.,
\begin{equation}
\label{eq:rank1constr}
  a_{ij} \ = \ 2ij - i^2 + j^2 \, , \qquad 
  b_{ij} \ = \ 2ij + i^2 - j^2 \, .
\end{equation}
Since $A + B = (4ij)_{i,j}$, the matrix $A+B$ is of rank~1.
Note that the game $(A,B)$ is \emph{symmetric}, i.e., $A = B^T$.

\begin{lemma} \label{le:twopureresponses}
For any mixed strategy $x \in \mathcal{S}_1$ there are at 
most two pure best responses for player~2.
And for any mixed strategy $y \in \mathcal{S}_2$ there are at 
most two pure best responses for player~1.
\end{lemma}

\begin{proof}
Let $y$ be a mixed strategy of player~2 with support 
$J := \{j_1, \ldots, j_k\}$.
We assume that there exists a 3-element subset
$I = \{i_1, i_2, i_3\} \subset \{1, \ldots, d\}$ such that
\begin{equation}
  \label{eq:inproofbestresponse}
  (Ay)_{i_1} \ = \ (Ay)_{i_2} \ = \ (Ay)_{i_3} \ \ge \ (Ay)_i
  \text{~ for all } i \not\in I \, .
\end{equation}
The equations in~\eqref{eq:inproofbestresponse} imply that for all 
distinct $i,i' \in I$ we have
\[
  \sum_{j \in J} \left( 2 i j - i^2 + j^2 \right) y_j \ = \ 
  \sum_{j \in J} \left( 2 i' j - i'^2 + j^2 \right) y_j \, ,
\]
which, using $\sum_{j \in J} y_j = 1$, is equivalent to
$
  2 (i - i') \sum_{j \in J} j y_j \ = \ (i^2 - i'^2) \, .
$
Hence,
  $2 \sum_{j \in J} j y_j \ = \ (i + i')$.
The left hand side of this equation is independent of $i$. Therefore
there cannot
be more than two indices in $I$ such that this equation is satisfied for all
pairs of these indices. 

The proof of the other statement is symmetric.
\end{proof}

\begin{lemma}
\label{le:vertices}
Each of the two polyhedra 
$\overline{P}$ and $\overline{Q}$ has $\frac{d}{6}(d^2+5)$ vertices,
which come in two classes:
\begin{enumerate}
\item There exists a $j \in \{1, \ldots, d\}$ such that the
  best response inequality of $\overline{Q}$ with index $j$ is binding and all
  nonnegativity 
  inequalities of $\overline{Q}$ but the one with index $j$ are binding  
  ($d$~vertices).
\item There exist $j_1,j_2 \in \{1, \ldots, d\}$, $j_1 < j_2$ 
  and $i \in \{j_1, \ldots, j_2-1\}$ such
  that the best response inequalities with indices $i$ and $i+1$
  are binding and all nonnegativity inequalities except those with indices
  $j_1,j_2$ are binding (altogether $\sum_{k=1}^{d-1} k (d-k)$ vertices).
\end{enumerate}
And similarly for $\overline{P}$.
\end{lemma}

\begin{proof}
We consider the polyhedron $\overline{Q}$.
By Lemma~\ref{le:twopureresponses},
at most two best response inequalities can be binding at a vertex
of~$\overline{Q}$.

If there is a single binding best response inequality, say, with
index $i$, then, at a vertex~$v$, at least $d-1$ of the nonnegativity 
inequalities
must be binding, and therefore there exists a single index $j$
such that $y_j$ is nonzero; hence $y_j = 1$. Now the condition
$v \in \overline{Q}$ implies $a_{ij} \ge a_{ij'}$ for all
$j' \in \{1, \ldots, d\}$, and it suffices to observe that
for a fixed $j$ the value $a_{ij}$ is maximized for $i = j$,
and this defines indeed a vertex.

Now assume that there are two binding best response inequalities
$i_1$ and $i_2$ with $i_1 < i_2$. 
Then there are at most two nonzero components of
$y$, say $y_{j_1}$ and $y_{j_2}$. We can assume that $j_1 \neq j_2$
since otherwise we are in the situation discussed before.

We claim that $i_1$ and $i_2$ are neighboring indices.
Otherwise there would exist an $i'$ with $i_1 < i' < i_2$.
Now, similar to the calculations in the proof of Lemma~\ref{le:twopureresponses}, 
the property $i' + i_2 > i_1 + i_2$ implies that
$2 (j_1 y_{j_1} + j_2 y_{j_2}) = (i_1+i_2) < (i'+i_2)$ and therefore
\[
  (Ay)_{i'} \ > \ (Ay)_{i_1} \ = \ (Ay)_{i_2} \, .
\]
This contradicts $v \in \overline{Q}$.

Now let $i_2 = i_1 + 1$. Computing the solutions for $y_{j_1}$
and $y_{j_2}$ of the equations
\begin{eqnarray*}
  2 j_1 y_{j_1} + 2 j_2 y_{j_2} & = & i_1 + i_2 \, , \\
  y_{j_1} + y_{j_2} & = & 1
\end{eqnarray*}
yields 
\[
  y_{j_1} \ = \ \frac{2 j_2 - (i_1+i_2)}{2 (j_2 - j_1)} \, ,
  \qquad
  y_{j_2} \ = \ \frac{(i_1+i_2) - 2 j_1}{2 (j_2 - j_1)} \, ,
\]
which in connection with $y \ge 0$ shows $j_1 \le i_1$ and
$j_2 > i_1$.

It remains to show that the stated pairs indeed define vertices.
In order to prove this, we have to show that for $i' < i_1$ or
$i' > i_2$ we obtain $(Ay)_{i'} < (Ay)_{i_1}$, which follows in
the same way as in the case $i_1 < i' < i_2$ that was discussed before.

Now summing up over all the possibilities proves the stated number.
\end{proof}

\begin{cor} \label{co:constrnasheq}
A pair of mixed strategies $(x,y)$ is a Nash equilibrium of the game $(A,B)$
if and only if $x=y=e_i$ for some unit vector $e_i$, $1 \le i \le d$, or
$x=y=\frac{1}{2}(e_i + e_{i+1})$ for some $i \in \{1, \ldots, d-1\}$.
\end{cor}

\begin{proof}
By the characterization of the vertices in Lemma~\ref{le:vertices},
the Nash equilibria come in two classes.
If for some $i \in \{1, \ldots, d\}$ both players play the
$i$-th pure strategy, then this gives a Nash equilibrium.
Moreover, for every $i \in \{1, \ldots, d-1\}$, if both players
only use the $i$-th and the $(i+1)$-th pure strategy, there exists
a Nash equilibrium. It is easy to check that in this situation,
both players play both of their pure strategies with 
probability~$\frac{1}{2}$. 
\end{proof}

Combining Theorem~\ref{th:linearbound} for rank 1-games
with von Stengel's result, we obtain the following lower bound
for rank $k$-games.

\begin{cor} \label{th:hierarchy}
For odd $d \ge 3$ and $k \le d$,
there exists a $d \times d$-game of
rank~k with at least $\tau(k-1) \cdot (2(d-k)+1)$ Nash equilibria,
where $\tau$ is defined as in~\eqref{eq:tau}.
For fixed $k$, this sequence converges to $\infty$ as $d$ tends
to $\infty$.
\end{cor}

\begin{proof}
We construct a $d \times d$-game $(A,B)$ of rank $k$ with
\[
  A \ = \ \left( \begin{array}{c|c}
     A' & 0 \\ \hline
     0  & A''
  \end{array} \right) \quad \text{ and } \quad
  \left( \begin{array}{c|c}
    B' & 0 \\ \hline
    0  & B''
  \end{array} \right)
\]
where $A',B' \in \R^{k-1} \times \R^{k-1}$ define a 
$(k-1) \times (k-1)$-game 
with $\tau(k-1)$ Nash equilibria, which exists by von Stengel's construction.
Moreover, let $A'',B'' \in \R^{d-k+1} \times \R^{d-k+1}$ 
define a $(d-k+1) \times (d-k+1)$-game
of rank~1 with $2(d-k+1)-1$ Nash equilibria based on 
the construction in Theorem~\ref{th:linearbound}.
Then the game $(A,B)$ is of rank $k$ and has at least
$\tau(k-1) \cdot (2(d-k)+1)$ Nash equilibria.
\end{proof}

\begin{rem}
Generalizing the construction in~\eqref{eq:abarbbar}, for a mapping 
$g: \{1, \ldots, d\} \to \R$ and a polynomial $p = \sum_{i=0}^n a_i x^i$ 
of degree $n$, the matrix $C \in \R^{d \times d}$ defined by
\[
  c_{ij} \ = \ p(g(i) - g(j))
\]
has rank at most $\frac{1}{2}(n+1)(n+2)$. This follows immediately
from applying the Binomial Theorem on $p(g(i) - g(j))$,
\[
  p(g(i) - g(j)) \ = \ 
  \sum_{k=0}^n a_k \sum_{l=0}^k \binom{k}{l} g(i)^l (-g(j))^{k-l} \, ,
\]
and observing that the rank of $C$ is bounded by the number of terms in this expansion.
\end{rem}

\section{Approximation algorithms}

\subsection{$\varepsilon$-approximating Nash equilibria of low rank games\label{se:approximation}}

For general bimatrix games, no polynomial time algorithm 
for $\varepsilon$-approximating a Nash equilibrium is known.
In a related model to ours, \cite{lmm-03} has provided the
first subexponential algorithm for finding an approximate
equilibrium.

Here, we show the following result for our restricted class
of bimatrix games.

\begin{thm} \label{th:approxlowranknash}
Let $k$ be a fixed constant and $\varepsilon > 0$. 
If $A+B$ is of rank $k$ then an $\varepsilon$-approximate 
Nash equilibrium can be found in time
$\mathrm{poly}(\mathcal{L},1/\varepsilon)$,
where $\mathcal{L}$ is the bit length of the input.
\end{thm}

Set 
\[
  Q \ = \ \left( \begin{array}{c|c}
    0 & \frac{1}{2}(A+B) \\ [2ex] \hline \\ 
    \frac{1}{2}(A^T + B^T) & 0
   \end{array} \right)
  \text{ \quad  and \quad } 
  z \ = \ \left( \begin{array}{c}x \\ y \end{array} \right)
\]
so that we the quadratic form $x^T (A+B) y$ can be written as 
$\frac{1}{2}z^T Q z$
with a symmetric matrix $Q$.
We assume that $A+B$ has rank $k$ for a fixed constant $k$; 
thus $Q$ has rank $2k$. Since the trace of the matrix $Q$ is
zero, this matrix is either the zero matrix or an indefinite matrix.
Hence, in the case $Q \neq 0$ the quadratic form defined by $Q$
is indefinite.

We use the following straightforward formulation of a Nash equilibrium
as a solution of a system of linear and quadratic inequalities.

\begin{lemma}
A pair of mixed strategies
$z = \binom{x}{y} \in \mathcal{S}_1 \times \mathcal{S}_2$ 
is a Nash equilibrium if and only if
there exists an $s \in \R$ such that
\begin{eqnarray*}
  z^T Q z & \ge & s \label{eq:s1} \\ 
 s & \ge & \left( A^{(i)} \, | \, B_{(j)}^T \right) z 
  \, \text{ \qquad for all 
  $i \in \{1, \ldots, m\}, \, j \in \{1, \ldots, n\}$.}
  \label{eq:s2}
\end{eqnarray*}
\end{lemma}
Since $z^T Q z \le s$ in any feasible solution of this optimization 
problem, we have $z^T Q z = s$ for any feasible solution. Hence,
the Nash equilibria are exactly the optimal solutions of the
quadratic optimization problem
\begin{equation}
  \label{eq:nashquadratic}
  \begin{array}{l@{\quad}l@{\qquad}rcl}
  \mathrm{(QP:)} & \multicolumn{4}{l}{\min s - z^T Q z} \\
  && s & \ge & \left( A^{(i)} \, | \, B_{(j)}^T \right) z
    \, \text{ \qquad for all 
    $i \in \{1, \ldots, m\}, \, j \in \{1, \ldots, n\}$,} \\
  && z & \in & \mathcal{S}_1 \times \mathcal{S}_2 \, .
  \end{array} 
\end{equation}

Vavasis has shown the following polynomial approximation 
result for quadratic optimization problems with compact polyhedral
feasible set \cite{vavasis-91,vavasis-92}.

\begin{prop} \label{pr:approxqp}
Let $\min \{\frac{1}{2}x^T Q x + q^T x \, : Ax \le b\}$ be a quadratic
optimization problem with compact support set $\{x \in \R^n \, : \, Ax \le b\}$, 
and let the rank $k$ of $Q$ be a fixed constant.
If $x^*$ and $x^{\#}$ denote points
minimizing and maximizing the objective function
$f(x) := \frac{1}{2}x^T Q x + q^T x$ in the feasible region, respectively,
then one can find in time $\mathrm{poly}(\mathcal{L},1/\varepsilon)$
a point $x^{\Diamond}$ satisfying
\[
  f(x^{\Diamond}) - f(x^*) \ \le \ \varepsilon(f(x^{\#}) - f(x^*)) \, ,
\]
where $\mathcal{L}$ is the bit length of the quadratic problem.
Such a point $x^{\Diamond}$
is called an \emph{$\varepsilon$-approximation} of the quadratic
problem.
\end{prop}

\smallskip

\noindent
{\sc Proof of Theorem~\ref{th:approxlowranknash}.}
The feasible region of the quadratic program~\eqref{eq:nashquadratic}
is unbounded. 
Since the value of $z^T Q z$ is at most $|A+B|$ for any feasible
solution $z$ and since the objective value for a Nash equilibrium is 0,
we can add the constraint $s \le |A + B|$ to~\eqref{eq:nashquadratic},
which makes the feasible region compact. Denote the resulting
quadratic optimization problem by $\textrm{QP'}$ and recall that
the approximation ratio of the quadratic program depends on 
the maximum objective value in the feasible region.

By Proposition~\ref{pr:approxqp}, we can compute in polynomial time 
an $\varepsilon$-approximation
$(z^{\Diamond},s^{\Diamond})$ with $z^{\Diamond} = 
(x^{\Diamond}, y^{\Diamond})$
of $\mathrm{QP'}$. Since the optimal value of $\mathrm{QP'}$ is $0$,
we have
\[
  s^{\Diamond} - (z^{\Diamond})^T Q z^{\Diamond} 
  \ = \ f(z^{\Diamond},s^{\Diamond}) 
  \ \le \ \varepsilon f(z^{\#},s^{\#})
  \ \le \ \varepsilon |A  + B| 
\, . 
\]
Hence, $(x^{\Diamond},y^{\Diamond})$ 
is an $\varepsilon$-approximate Nash equilibrium 
of the game $(A,B)$.
\hfill $\Box$

\begin{rem} 
The proof in \cite{vavasis-91} computes an
$LDL^T$ factorization of the matrix $Q$ defining the quadratic form
and then constructs a sufficiently fine grid in the fixed-dimensional space.
Since the quadratic form $x^T Q y$ is bilinear, we can also directly apply 
an $LDU^T$ factorization on the matrix of the bilinear form.
\end{rem}

\subsection{Relative approximation in case of a nonnegative 
decomposition\label{se:relativeapprox}}

The right hand side in Definition~\ref{de:approxnasheq} of an
approximate Nash equilibrium depends only on $\varepsilon$ and
on $|A+B|$. Since different Nash equilibria in the same
game can differ strongly in their payoffs, we introduce a notion
of \emph{relative approximation} with respect to a Nash payoff
which takes into account these differences.

Consider the quadratic problem~\eqref{eq:nashquadratic}.
In a Nash equilibrium $(x,y) \in \mathcal{S}_1 \times \mathcal{S}_2$
there exists an $s \in \R$ such that $(x,y,s)$ is a feasible solution
to~\eqref{eq:nashquadratic}; in this situation $s$ coincides with
the sum of the payoffs of the two players. 
In the relative approximation, we aim
at finding pairs of strategies $(x,y)$ for which 
there exists an $s \in \R$ such that $(x,y,s)$ is feasible and
\[
  s - x^T (A+B) y \ \le \ \rho s \, .
\]
Using our notion of loss, by observing 
$s = \max_i A^{(i)} x + \max_j x^T B_{(j)}$ for an optimally chosen $s$, 
this means
\[
  \ell(x,y) \ \le \ \rho (\max_i A^{(i)} x + \max_j x^T B_{(j)}) \, .
\]

We provide an efficient approximation algorithm for the case that
$C = A+B$ has a known decomposition of the form
\begin{equation}
  \label{eq:nonnegdecomp}
  C \ = \ \sum_{i=1}^k u^{(i)} (v^{(i)})^T
\end{equation}
with non-negative vectors $u^{(i)}$ and $v^{(i)}$. 

\begin{thm} \label{th:relativeapprox}
If $C$ has a known decomposition of the 
form~\eqref{eq:nonnegdecomp} then for any given $\varepsilon > 0$
a relatively approximate Nash equilibrium with approximation ratio
$1-\frac{1}{(1 + \varepsilon)^2}$ can be computed in time
$\mathrm{poly}(\mathcal{L},1 / \log(1 + \varepsilon))$, where
$\mathcal{L}$ is the bit length of the input.
\end{thm}

Let $z_i = x^T \cdot u^{(i)}$,
$w_i = (v^{(i)})^T \cdot y$.
We put a grid on each of the $z_i$ and on each of the $w_i$
in a geometric progression: denoting by
\[
  (z_i)_{\min} \ = \ \min_{x \in \mathcal{S}_1} x^T \cdot u^{(i)}
  \qquad \text{ and }
  (z_i)_{\max} \ = \ \max_{x \in \mathcal{S}_1} x^T \cdot u^{(i)}
\]
the minimum and the maximum possible value for $z_i$,
we partition the interval $[(z_i)_{\min}, (z_i)_{\max}]$
into the intervals
$[(z_i)_{\min}, (1 + \varepsilon) (z_i)_{\min}]$,
$[(1 + \varepsilon)(z_i)_{\min}, (1 + \varepsilon)^2 (z_i)_{\min}]$,
and so on. And analogously for the $w_i$.

For every cell we construct a linear program which 
``approximates'' the quadratic program~\eqref{eq:nashquadratic}.
Let the intervals of
a grid cell be $[\alpha_i, (1 + \varepsilon) \alpha_i]$ and 
$[\beta_i, (1 + \varepsilon) \beta_i]$, i.e.,
\[
  \begin{array}{ccccc}
  \alpha_i & \le & z_i & \le & (1+\varepsilon) \alpha_i \, , \\ [1ex]
  \beta_i & \le & w_i & \le & (1+\varepsilon) \beta_i \, . \\
  \end{array}
\]
Then for any pair of strategies $(x,y) \in \mathcal{S}_1 \times \mathcal{S}_2$
falling into that cell, the quadratic form $x^T C y$ satisfies
\begin{equation}
  \label{eq:xcyest}
  \sum_{i=1}^k \alpha_i \beta_i \ \le \ x^T C y \ \le \  
  (1 + \varepsilon)^2
  \sum_{i=1}^k \alpha_i \beta_i \, ,
\end{equation}
where the left inequality uses that all the values in the decomposition
are nonnegative. For the grid cell, we consider the linear program
\[
  \begin{array}{ccccl}
    \multicolumn{3}{l}{\min s - \sum_{i=1}^k \alpha_i \beta_i} \\
    \alpha_i & \le &  x^T \cdot u^{(i)} & \le & (1 + \varepsilon) \alpha_i \, , \\
    \beta_i & \le & (v^{(i)})^T \cdot y & \le & (1+\varepsilon) \beta_i \, , \\
     s & \ge & \multicolumn{3}{l}{\left( A^{(i)} \, | \, B_{(j)}^T \right) z
    \, \text{ \qquad for all 
    $i \in \{1, \ldots, m\}, \, j \in \{1, \ldots, n\}$,}} \\
    (x,y) & \in & \multicolumn{3}{l}{\mathcal{S}_1 \times \mathcal{S}_2 \, , \; s \in \R \, .}
  \end{array}
\]
In at least one of the cells there exists a Nash equilibrium.
The linear program corresponding to that cell yields a solution
with
\begin{equation}
\label{eq:nashest}
  \sum_{i=1}^k \alpha_i \beta_i 
  \ \le \ s \ \le \ 
  (1+\varepsilon)^2 
  \left( \sum_{i=1}^k \alpha_i \beta_i \right) \, .
\end{equation}
Hence, by the left inequality in~\eqref{eq:xcyest} and the right
inequality in~\eqref{eq:nashest} we have
\[
  x^T C y 
              \ \ge \ \sum_{i=1}^k \alpha_i \beta_i 
              \ \ge \ \frac{s}{(1+\varepsilon)^2} \, .
\]
We conclude
\[
  s - x^T C y \ \le \ s \left( 1 - \frac{1}{(1+\varepsilon)^2} \right) \, ,
\]
which shows Theorem~\ref{th:relativeapprox}.

\providecommand{\bysame}{\leavevmode\hbox to3em{\hrulefill}\thinspace}
\providecommand{\MR}{\relax\ifhmode\unskip\space\fi MR }
\providecommand{\MRhref}[2]{%
  \href{http://www.ams.org/mathscinet-getitem?mr=#1}{#2}
}
\providecommand{\href}[2]{#2}


\end{document}